\documentclass[%
prl%
 ,twocolumn%
 ,superscriptaddress%
,tightenlines%
,amssymb,amsmath,nobibnotes, aps, prl]{revtex4}
\usepackage{dcolumn}
\usepackage{bm}
\usepackage[dvips]{graphicx}
\begin{document}
\title{Spin Degree of Freedom in the $\nu =1$ Bilayer Electron System\\Investigated via Nuclear Spin Relaxation}
\author{N. Kumada}
\author{K. Muraki}
\affiliation{NTT\,Basic\,Research\,Laboratories,\,NTT\,Corporation,\,3-1\,Morinosato-Wakamiya,\,Atsugi\,243-0198,\,Japan}
\author{K. Hashimoto}
\altaffiliation[Present address: ]{Institute of Applied Physics, Hamburg University, Jungiusstrasse 11, D-20355 Hamburg, Germany.}
\affiliation{SORST-JST, 4-1-8 Honmachi, Kawaguchi, Saitama 331-0012, Japan}
\author{Y. Hirayama}
\affiliation{NTT\,Basic\,Research\,Laboratories,\,NTT\,Corporation,\,3-1\,Morinosato-Wakamiya,\,Atsugi\,243-0198,\,Japan}
\affiliation{SORST-JST, 4-1-8 Honmachi, Kawaguchi, Saitama 331-0012, Japan}

\date{Version: \today}

\begin{abstract}
The nuclear-spin-relaxation rate $1/T_1$ has been measured in a bilayer electron system at and around total Landau level filling factor $\nu =1$.
The measured $1/T_1$, which probes electron spin fluctuations, is found to increase gradually from the quantum Hall (QH) state at low fields through a phase transition to the compressible state at high fields.
Furthermore, $1/T_1$ in the QH state shows a small but noticeable increase away from $\nu =1$.
These results demonstrate that, as opposed to common assumption, the electron spin degree of freedom is completely frozen neither in the QH nor compressible states.
\end{abstract}
\maketitle


The bilayer electron system at total Landau level filling factor $\nu =1$ ($1/2$ in each layer) have been continuously drawing intensive research interest because of its unique phase diagram \cite{Murphy,Sarmabook}.
Two parameters play crucial role: the ratio of the intralayer to interlayer Coulomb interactions, given by the ratio between the interlayer distance $d$ and the magnetic length $l_B=\sqrt{\hbar /eB}$ (and hence controlled by the magnetic field $B$), and the tunneling gap $\Delta _{\rm SAS}$. 
For $d/l_B$ below a certain critical value, strong interlayer interactions lead to a many-body quantum Hall (QH) state even in the limit of zero tunneling gap, supported by spontaneous interlayer coherence \cite{Spielman}.
When intralayer interactions dominate at larger $d/l_B$, in contrast, the QH state collapses into a compressible state having lower intralayer correlation energy even for finite $\Delta _{\rm SAS}$ \cite{MacDonald}.
Recent interests have focused on the nature of the phase transition between these limits.
Interlayer tunneling \cite{Spielman} and Coulomb drag \cite{Kellogg2003} experiments carried out on samples with vanishingly small $\Delta _{\rm SAS}$ have revealed that below a critical $d/l_B$ the interlayer coherence develops somewhat continuously, in contrast to the result of a numerical study \cite{Schliemann2001} that the transition is likely to be first order at any value of $\Delta _{\rm SAS}$.
In spite of various theoretical models \cite{Nomura,Stern,Simon,Sheng} motivated by these findings, a comprehensive understanding is not yet achieved.

In all these theories, it is routinely assumed that the spin degree of freedom is frozen by the Zeeman coupling to the magnetic field, and physics is governed solely by the pseudospin representing the layer degree of freedom.
This assumption seems reasonable for the QH state, and is indeed verified by numerical calculation \cite{Schliemann2001}.
We note, however, that because of the small $g$-factor in GaAs, a monolayer two-dimensional electron system (2DES) at $\nu =1/2$ is not fully spin polarized in the relevant magnetic field region, $B\lesssim 10$\,T \cite{Kukushkin, Dementyev,Melinte2000}.
Hence, if the bilayer $\nu =1$ compressible state consists of two independent monolayers of $1/2$ fillings, the phase transition must be accompanied by changes not only in the pseudospin but also in the spin configurations.
While spin states in monolayer systems have been studied by various optical means \cite{Kukushkin}, little is known about the spin states in bilayer systems, presumably because of the difficulty in independently controlling the electron density in each layer.

In this Letter, we investigate the electron spin degree of freedom in a bilayer system at total filling $\nu =1$ via nuclear spin relaxation of the host GaAs.
Since the Zeeman energy of nuclear spin is about three orders of magnitude smaller than that of electron spin, exchange of spin angular momentum between the two spin subsystems is allowed only when the electron system supports spin excitations with zero energy.
The nuclear spin relaxation rate $1/T_1$ thus probes the spectral density at zero energy of the electron spin system.
While nuclear spins have often been used as a useful probe of the electron spin state, a small number of nuclei in contact with the 2DES and the overwhelming background due to the thick substrate have limited their applicability mostly to multilayer samples with fixed electron densities \cite{Melinte2000}.
We here employed a current-pump and resistive-detection technique \cite{Hashimoto}, which enables us to measure $1/T_1$ in a single pair of fully-density-tunable 2DESs over a wide range of magnetic field.
As $d/l_B$ is increased from the QH to compressible states, we observe that $1/T_1$ shows a smooth increase across the phase boundary, rather than a sudden jump expected for a first order transition.
Furthermore, $1/T_1$ in the QH state shows a small but noticeable increase away from $\nu =1$.
These results demonstrate that, as opposed to common assumption, the electron spin degree of freedom is completely frozen neither in the QH nor compressible states.

The sample used in this study consists of two 200-\AA-wide GaAs quantum wells separated by a 15-\AA-thick AlAs barrier (center-to-center distance of $d=215$\,\AA), processed into a 50-$\mu $m-wide Hall bar with ohmic contacts connecting to both layers.
By adjusting the front- and back-gate biases, we control the filling factors in the front layer $\nu _f$ and back layer $\nu _b$ independently.
The low temperature mobility is 2.0$\times 10^{6}$\,cm$^2/$V\,s for a total density of $n_t=2.0\times 10^{11}$\,cm$^{-2}$.
From low-field Shubnikov-de Haas oscillations $\Delta _{\rm SAS}$ is estimated to be 4\,K, in reasonable agreement with calculation.
For this sample, the QH-compressible phase transition occurs at $B=8.5$\,T ($d/l_B=2.45$).
Measurements of the longitudinal resistance $R_{xx}$, and hence $1/T_1$ (see below), were performed at $T=0.3$\,K, using a standard low-frequency lock-in technique.
A relatively large current of 100\,nA was used to induce nuclear polarization \cite{Kraus}.

\begin{figure}[t]
\begin{center}
\includegraphics[width=0.88\linewidth]{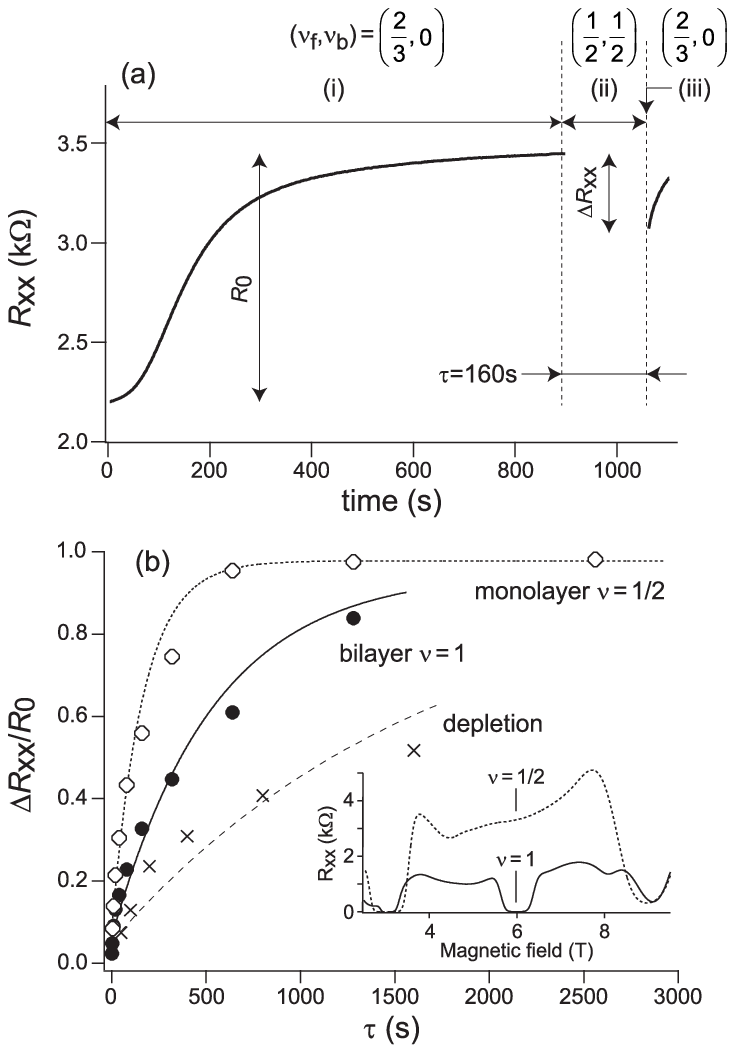}
\caption{$1/T_1$ measurements for $\nu =1$ in the balanced bilayer at $B=6$\,T.
(a) An example of resistively-detected nuclear spin relaxation for $\tau =160$\,s.
(b) $\tau $ dependence of $\Delta R_{xx}/R_0$ (solid circle).
Data for $\nu =1/2$ in the monolayer (open circle) and depletion (cross) are also shown.
Lines are fitted curves.
The inset shows $R_{xx}$ for the balanced bilayer at $n_t=1.46\times 10^{11}$\,cm$^{-2}$ (solid) and monolayer at $n_t=7.3\times 10^{10}$\,cm$^{-2}$ (dashed).
}
\label{sequence}
\end{center}
\end{figure}

The current-pump and resistive-detection technique used here was originally developed for a monolayer system \cite{Hashimoto}; it exploits the current-induced nuclear spin polarization and resultant change in $R_{xx}$ that occur in a monolayer system at $\nu \simeq 2/3$ in the vicinity of its QH ferromagnetic transition point \cite{Kraus}.
The procedure of the measurements is as follows [Fig.\,\ref{sequence}(a)].
First, nuclear spin polarization is current-induced in the front layer by setting $\nu _f\simeq 2/3$ and $\nu _b=0$, where $R_{xx}$ slowly increases with a typical time constant of 10\,min and saturates at a value enhanced by an amount $R_0$ from the initial value [step (i)].
Then, using the gates the filling factors are quickly changed and kept at temporary values for a given period of time $\tau$, during which the nuclear spins are allowed to relax towards equilibrium [step (ii)].
Subsequently, the filling factors are restored to $\nu _f\simeq 2/3$ and $\nu _b=0$, where we read the resistance relaxation $\Delta R_{xx}$ [step (iii)], which provides a measure of the nuclear spin relaxation during $\tau $ \cite{Hashimoto}.
In the following, the filling factors referred to are exclusively those set during the relaxation procedure of step (ii).

Figure\,\ref{sequence}(a) depicts a measurement carried out at $B=6$\,T with $\tau =160$\,s where we chose $\nu _f=\nu _b=1/2$, i.e., $\nu =1$ in the balanced bilayer.
At this magnetic field, $d/l_B=1.9$ ($<2.45$) and hence the system is deep in the QH regime as shown by the well developed $R_{xx}$ minimum [inset to Fig.\,\ref{sequence}(b)].
By repeating similar measurements for different values of $\tau $ [solid circles in Fig.\,\ref{sequence}(b)] and fitting the data to the relation $\Delta R_{xx}/R_0=1-\exp (-\tau /T_1)$, we derive a relaxation rate, $1/T_1=1.9\times 10^{-3}$\,s$^{-1}$.
This value is to be compared with $1/T_1=6.1\times 10^{-3}$\,s$^{-1}$ obtained for the monolayer system at $\nu =1/2$ ($\nu _f=1/2$ and $\nu _b=0$) (open circles).
Since in these two cases the density in the layer where $1/T_1$ is measured (front layer) is identical, the difference in $1/T_1$ reflects the difference of electronic states \cite{current}.
Since $\nu =1/2$ corresponds to the zero effective magnetic field for composite fermions (CFs) \cite{HLR}, the larger $1/T_1$ for the $\nu =1/2$ monolayer system can be understood as due to the continuous density of states (DOS) and the coexistence of opposite spins at the Fermi level of CFs.
This allows for spin-flip scatterings with zero energy, thereby enhancing $1/T_1$ \cite{Dementyev,Hashimoto}.
The smaller $1/T_1$ for the bilayer $\nu =1$ QH state, in turn, implies that such spin-flip scatterings are suppressed.
In this state, all electrons occupy the spin-up state of the lowest Landau level in the symmetric subband, and hence the spin excitation mode has a gap equal to the Zeeman energy \cite{Burkov}.
The spin angular momentum exchange between electronic and nuclear subsystems is therefore inhibited.
We however note that the $1/T_1$ for the bilayer $\nu =1$ QH state is noticeably larger than the value, $1/T_1=5.9\times 10^{-4}$\,s$^{-1}$, for the case of complete depletion, $\nu _f=\nu _b=0$ (crosses), which is considered to be due to nuclear spin diffusion.
This indicates that there is a finite contribution from the 2DES to $1/T_1$ even in the QH state.

\begin{figure}[t]
\begin{center}
\includegraphics[width=0.85\linewidth]{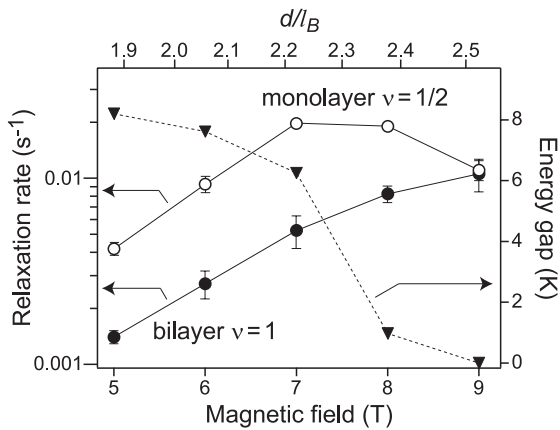}
\caption{$1/T_1$ for the $\nu =1$ bilayer system (closed circle) and $\nu =1/2$ monolayer system (open circle) as a function of magnetic field.
$d/l_B$ is shown on the top axis.
The energy gap (triangle) of the bilayer $\nu =1$ QH state is also plotted.
}
\label{extract}
\end{center}
\end{figure}

To investigate the QH-compressible phase transition, we performed similar measurements for several magnetic fields.
Results are shown in Fig.\,\ref{extract} \cite{cooldown}, together with the energy gap of the bilayer $\nu =1$ QH state determined separately from the temperature dependence of the $R_{xx}$ minimum.
It is seen that $1/T_1$ for the $\nu =1/2$ monolayer system grows with $B$ and declines above 7\,T.
The decrease at high fields signals that the spin polarization is about to become complete at these fields, consistent with previous reports \cite{Kukushkin,Dementyev,Melinte2000}.
The increase at lower fields is thought to reflect the DOS at the Fermi level of CFs \cite{DOS}.
The most important observation here, however, is that $1/T_1$ for the $\nu =1$ bilayer system steadily increases with $B$ over the whole magnetic field range until it becomes nearly equal to that of the $\nu =1/2$ monolayer system at $B=9$\,T.
It is to be noted that this magnetic field coincides with the field at which the energy gap of the bilayer $\nu =1$ QH state collapses.

The observed increase in $1/T_1$ with $B$ in the QH state is rather unexpected.
As already mentioned, in the presence of a magnetic field the spin excitation mode of a fully polarized ground state has a finite gap equals to the Zeeman energy, and therefore does not couple with nuclear spins.
Moreover, the spin gap increases in proportion to $B$.
Although intralayer interactions ($\propto B^{1/2}$) soften the pseudospin mode and eventually destroys the QH state at high fields \cite{MacDonald}, their influence on the spin mode is very minor \cite{Burkov}.
Hence, one can rule out thermally excited spin waves as the cause of the field dependence of $1/T_1$ in the QH state.

\begin{figure}[t]
\begin{center}
\includegraphics[width=1.0\linewidth]{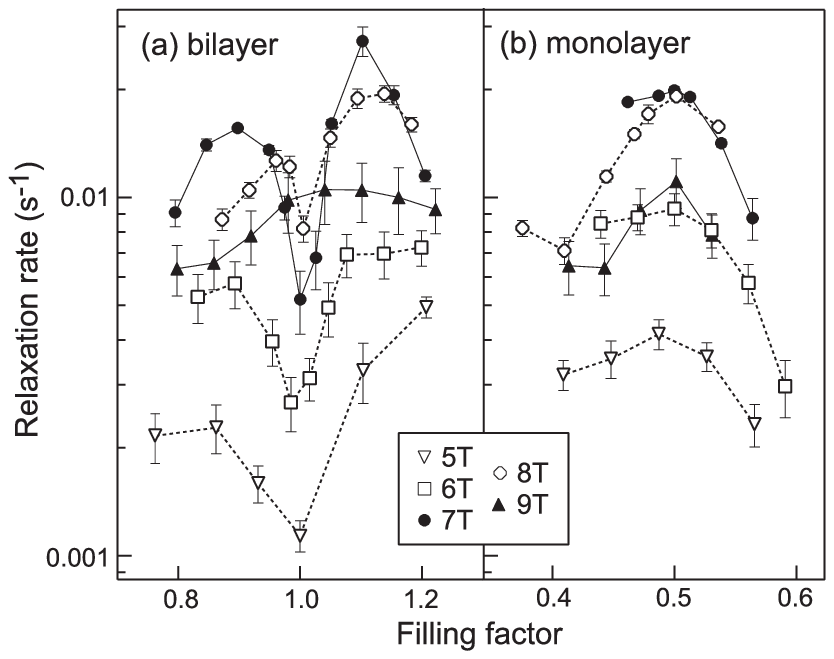}
\caption{$1/T_1$ around (a) $\nu =1$ in the balanced bilayer and (b) $\nu =1/2$ in the monolayer.
}
\label{sigma0}
\end{center}
\end{figure}

Further insight is gained by investigating $1/T_1$ while varying the filling around $\nu =1$.
Figure\,\ref{sigma0}(a) shows data taken at several magnetic fields.
One finds that $1/T_1$ shows a clear minimum at $\nu =1$ at low fields where the QH state is well developed.
This in turn implies that the spin fluctuation increases as charged quasiparticles are introduced in the QH state.
This is not trivial, because the lowest-energy charged excitation in the bilayer $\nu =1$ QH state is believed to be associated with the pseudospin degree of freedom \cite{Sarmabook}, not the spin degree of freedom.

As the magnetic field is increased, the minimum in $1/T_1$ becomes smaller at 8\,T and eventually collapses at 9\,T, leaving a broad maximum.
This demonstrates that $1/T_1$ is a sensitive indicator of the QH-compressible transition.
This maximum is similar to the one observed for the monolayer system around $\nu =1/2$ [Fig.\,\ref{sigma0}(b)].
The latter can be naturally explained by the CF picture; away from $\nu =1/2$, the energy spectrum is quantized by the effective magnetic field experienced by CFs, and therefore spin-flip scattering with zero-energy is restricted.
The similarity between these maxima together with the quantitative agreement between the values of $1/T_1$ in the two systems at 9\,T (Fig.\,\ref{extract}) strongly suggest an intimate link between the bilayer $\nu =1$ compressible state and the $\nu =1/2$ CF metal.

We now discuss in more detail the $1/T_1$ data for the $\nu =1$ bilayer system.
The gradual change in $1/T_1$ in the vicinity of the phase transition may seem inconsistent with the first order phase transition anticipated in Ref.\,\cite{Schliemann2001}.
This, however, can be resolved if the QH state contains compressible puddles, which consist of charged quasiparticles, embedded in the incompressible background, as suggested in Ref.\,\cite{Stern}.
Such compressible puddles arise from long-range disorder due to remote donors, which causes local filling to deviate from $\nu =1$.
The fraction of the compressible regions should depend on the strength of the disorder and the energy gap for creating charged quasiparticles.
Hence, as $d/l_B$ increases and the energy gap becomes small, the compressible puddles grow until percolation takes place at a certain point, where the apparent gap collapses.
Since $1/T_1$ away from $\nu =1$ is as large as $1/T_1$ for the CF metals [Figs.\,\ref{sigma0}(a) and (b)], this explains the smooth evolution of $1/T_1$ in the vicinity of the phase transition.
This scenario can also account for the observation that the 2DES makes a finite contribution to $1/T_1$ even for sufficiently small $d/l_B$ where the QH state is well developed.
Indeed, for $B\leq 7$\,T, where the energy gap is only weakly dependent on $B$, $1/T_1$ for the QH state is seen to follow the behavior of $1/T_1$ for the CF metal (Fig.\,\ref{extract}).

\begin{figure}[t]
\begin{center}
\includegraphics[width=0.7\linewidth]{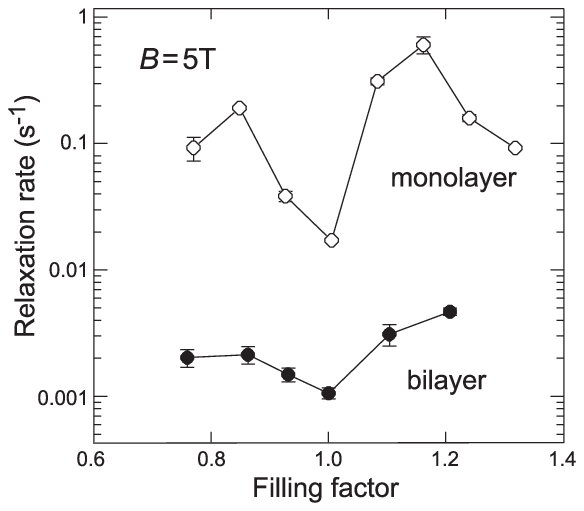}
\caption{$1/T_1$ around $\nu =1$ for the balanced bilayer (closed circle) and the monolayer (open circle).
}
\label{nu1}
\end{center}
\end{figure}

As a test, we compare in Fig.\,\ref{nu1} the behavior of $1/T_1$ around $\nu =1$ for the balanced bilayer ($\nu _f=\nu _b$) with that for the monolayer with all electrons transferred to the front layer ($\nu _b=0$).
$1/T_1$ for the monolayer is characterized by a drastic enhancement on both sides of $\nu =1$, due to the gapless spin mode that occurs in the presence of non-collinear spin alignment associated with skyrmions \cite{Tyckocote}.
$1/T_1$ for the bilayer is two orders of magnitude smaller, indicating that the role of spin is much smaller in the bilayer, and the quasiparticle is associated predominantly with the pseudospin degree of freedom.
We note that even at exact $\nu =1$, $1/T_1$ is much larger for the monolayer than for the bilayer, indicating that skyrmions are present even at exact $\nu =1$.
It is therefore reasonable to assume that the bilayer system at exact $\nu =1$ contains localized quasiparticles, or isolated compressible puddles, which then contribute to $1/T_1$.

Finally, we discuss two important issues left unaddressed.
The first is the possible role of finite tunneling in our sample.
According to Ref.\,\cite{Schliemann2001}, the order of the phase transition does not depend on $\Delta _{\rm SAS}$.
Hence, essentially the same arguments hold for samples with smaller $\Delta _{\rm SAS}$.
For such samples, the phase transition occurs at lower magnetic fields \cite{Spielman,Kellogg2003}.
Thus, the smaller Zeeman energy will make the role of spin even more important.
The lower electron densities and the smaller energy gap for such samples will also enhance the role of disorder.
Second, and most importantly, it remains unexplained why the quasiparticles associated predominantly with the pseudospin degree of freedom contribute to the nuclear spin relaxation.
There are theories connecting the bilayer $\nu =1$ QH state with CF liquids \cite{Simon,Veillette}.
Since the spin degree of freedom is not frozen in CF liquids, these theories may provide an important clue.
Another possibility may be an excitation including both spin and pseudospin \cite{EzawaSU4}.
Further discussion on this issue, however, is beyond the scope of this Letter.

To summarize, we have studied the electron spin degree of freedom of a bilayer $\nu =1$ system using resistively-detected nuclear spin relaxation.
The measured $1/T_1$ demonstrates that, as opposed to common assumption, the electron spin degree of freedom is completely frozen neither in the QH nor compressible states.
The behavior of $1/T_1$ in the QH state reveals that the quasiparticles are responsible for the nuclear spin relaxation, the mechanism of which remains to be explained.

The authors are grateful to T. Saku for growing the heterostructures, J. Schliemann, K. Nomura, K. Takashina, A. Sawada, and Z. F. Ezawa for discussions.

\bibliography{D}

\begin{thebibliography}{23}
\bibitem[{cur()}]{current}
\bibinfo{note}{We confirmed that effects of current during the relaxation
  procedure are negligible.}
\bibitem[{coo()}]{cooldown}
\bibinfo{note}{The data in Fig.\,\ref{extract} were obtained on a different
  cool down from the data in Fig\,\ref{sequence}.}
\bibitem[{DOS()}]{DOS}
\bibinfo{note}{The dependence on $B$ is found to be stronger than expected from
  a simple picture, and is not understood yet.}
\end{thebibliography}

\end{document}